\documentclass[twocolumn,aps,pre,showpacs,preprintnumbers,amsmath,amssymb]{revtex4}

\usepackage{graphicx}
\usepackage{dcolumn}
\usepackage{bm}
\begin{document}

{\parbox[b]{1in}{\hbox{\tt INT-PUB-09-034}}}
\title{The Structure of Physical Crystalline Membranes within the Self-Consistent Screening Approximation.}
\author{Doron Gazit}
\email{doron.gazit@mail.huji.ac.il}
\affiliation{Institute for Nuclear Theory, University of Washington, 
Box 351550, Seattle, WA 98195, USA}

\date{\today}

\begin{abstract}
The anomalous exponents governing the long wavelength behavior of the flat phase of physical crystalline membranes are calculated within a self-consistent screening approximation (SCSA) applied to second order expansion in $1/d_C$ ($d_C$ is the co-dimension), extending the seminal work of Le Doussal and Radzihovsky [Phys. Rev. Lett. {\bf 69}, 1209 (1992)]. In particular, the bending rigidity is found to harden algebraically in the long wavelength limit with an exponent $\eta=0.789...$, which is used to extract the elasticity softening exponent $\eta_u=0.422...$, and the roughness exponent $\zeta=0.605...$. 
The scaling relation $\eta_u=2-2\eta$ is proven to hold to all orders in SCSA. Further, applying the SCSA to an expansion in $1/d_C$, is found to be essential, as no solution to the self-consistent equations is found in a two bubble level, which is the na{\"\i}ve second order expansion. Surprisingly, even though the expansion parameter for physical membrane is $1/d_C=1$, the SCSA applied to second order expansion deviates only slightly from the first order, increasing $\zeta$ by mere $0.016$. This supports the high quality of the SCSA for physical crystalline membranes, as well as improves the comparison to experiments and numerical simulations of these systems. The prediction of SCSA applied to first order expansion for the Poisson ratio is shown to be exact to all orders.
\end{abstract}

\pacs{82.45.Mp, 87.16.D-, 61.46.-w, 11.15.Pg.}

\maketitle
\begin{section}{Introduction}
Physical membranes are 2-dimensional (2D) surfaces embedded in the
3-dimensional space. A subset of these are crystalline
membranes, known also as tethered or polymerized, that have a solidlike
structure of 2D lattice, usually triangular or hexagonal, with fixed connectivity \cite{Membranes_Book}. 

Examples of such systems are plentiful in our world. In biological systems, prominent is the intercellular side of the plasma membrane of some
types of cells, which is built of spectrin proteins in a triangular
lattice. Different examples, in the soft condensed matter field, are monolayers of  polymerized phospholipid molecules suspended on an air-water interface, using their amphiphilic nature \cite{Membranes_Book}. Recently, graphene, a single layer plane
of carbon atoms, as well as individual crystal planes of other layered materials, were isolated experimentally \cite{2004SciNovoselov,2005PNASNovoselov}, representing the ultimate crystalline membrane. Further studies have demonstrated the stability of graphene even when it is free standing, i.e, without the support of a substrate, and tensionless \cite{2007NatMeyer}.

The stability at finite temperature of a flat phase of tensionless crystalline membranes may seem like a violation of the Mermin-Wagner theorem, which forbids the
existence of long range order in 2D systems due to diverging thermal
vibrations. However, this seeming contradiction between experiment and theory
is resolved by introducing out-of-plane fluctuations. These fluctuations induce frustration between
the large thermal vibrations in 2D and the competing gain in elastic
energy, which stabilizes the globally flat phase even at finite temperatures \cite{Membranes_Book,1987JPhNelson,1988PRLAronovitz, 1988PhysRevAKardar,1988EPLDavid_1,1988EPLDavid_2, 1989JPhAronovitz, 1992PRLSCSA,1995PhysRevLettRadzihovsky, 1998PhysRevERadzihovsky,2009PhysRevEKownacki}.

This asymptotically flat phase is perturbed by the transverse displacements, whose amplitude
diverges with the size of the system $L$, as $L^\zeta$, where $\zeta$ is roughness exponent. This behavior is a result of the anomalous
bending energy of the flat phase, that for small wavevectors $q$ deviates from its constant value and acquires an anomalous exponent
$\kappa_R(q) \sim q^{-\eta}$, satisfying a scaling relation
$\eta=(4-D)-2\zeta$, where $D$ is the dimensionality of the surface,
i.e. $D=2$ for physical membranes. 
This picture, as well as other properties of the flat phase, is a conclusion of more than two decades of extensive theoretical, experimental and numerical research \cite{Membranes_Book,1987JPhNelson,1988PRLAronovitz, 1988PhysRevAKardar,1988EPLDavid_1,1988EPLDavid_2, 1989JPhAronovitz, 1992PRLSCSA,1993PhysRevEZhang,1993SciSchmidt,1995PhysRevLettRadzihovsky, 1998PhysRevERadzihovsky,1996PhysRevEZhang,1996JPhBowick,2009PhysRevEKownacki,2009arXivLos}. In their seminal
work, Nelson and Peliti \cite{1987JPhNelson} have estimated these
exponents to be $\eta=1,\,\zeta=\frac{1}{2}$, using a one-loop
expansion of the bending energy, within a self-consistent
approximation, and assuming a finite renormalization of the elastic
constants. 

Aronovitz and Lubensky \cite{1988PRLAronovitz} have used an epsilon
expansion $\epsilon=4-D$ to falsify this assumption, showing that the
elastic constants obtain an anomalous exponent $\eta_u$, vanishing at long distances as
$q^{\eta_u}$. Using Ward identities of the rotational group, they achieved a second
scaling relation $\eta_u=(4-D)-2\eta$.

The theoretical evaluation of the anomalous exponents which is considered to be the most accurate is within a self-consistent screening approximation
(SCSA), introduced in the context of crystalline membranes by Le
Doussal and Radzihovsky \cite{1992PRLSCSA}. This approximation is found
to reproduce known theoretical limits, viz. the limit of large embedding dimension $d$ \cite{1988EPLDavid_1,1988EPLDavid_2}; the case where the
surface dimension equals that of the embedding space; and the leading order $\epsilon$ expansion. 

However, when extrapolated to physical membranes, the exponents calculated in the different methods vary substantially. Extrapolating the large-$d$ expansion leads to $\eta=\eta_u=\zeta=2/3$; extrapolating the $\epsilon$ expansion predicts $\eta=0.96$, $\eta_u=0.08$ and $\zeta=0.52$; whereas SCSA predicts $\eta=\frac{4}{1+\sqrt{15}}=0.821...$, $\eta_u=0.358...$, and $\zeta=0.590...$ . A different value, though rather close to SCSA, was recently calculated by Kownacki and Mouhanna \cite{2009PhysRevEKownacki} that have applied a non-perturbative renormalization group (RG) approach to the problem of crystalline membranes, using a simple ansatz for the action, to lowest powers in derivative and field expansion, to predict $\eta=0.849$. 

Thus, it is important to assess the issue of accuracy of the SCSA prediction for physical membranes, in order to determine its relevance to this case. However, the SCSA is an uncontrolled approximation, as it includes a partial summation of infinite amount of diagrams. Thus, its uncertainty is unknown. The main goal of the current paper is to study the SCSA expansion and its accuracy by considering the effect of higher orders of the expansion. 

To date, going beyond the aforementioned approximations, and ``select" the theoretical approach which gives the most accurate results, demanded the
use of numerical simulations or experimental results. One can extract $\eta_u=0.50(1)$, $\zeta=0.64(2)$, and $\eta=0.750(5)$ from the Monte-Carlo simulations of Ref.~\cite{1996JPhBowick}, which should be compared with $\eta=0.81$ and $\zeta=0.59(2)$, found numerically in Ref.~\cite{1993PhysRevEZhang} and \cite{1996PhysRevEZhang}, respectively. Lately \cite{2009arXivLos}, the anomalous exponents were calculated for a sheet of carbon atoms (representing graphene without the electronic degrees of freedom), using a realistic carbon-carbon potential, giving $\eta\approx0.85$. 
These results  are compatible, though not perfectly, with the theoretical approaches of SCSA, non perturbative RG and large-d estimates, as well as with the existing experimental measurement of the static structure factor of the red blood cell cytoskeleton by small-angle x-ray and light scattering, yielding a roughness exponent $\zeta = 0.65(10)$ \cite{1993SciSchmidt}.

The paper is built as follows. After a description of the model and its long-wavelength properties, it is shown that the scaling relation $\eta_u =2-2\eta$ is exact to all orders of SCSA in D=2. Then, the second order SCSA is developed. Explicitly, it is demonstrated that a two loop expansion, which is a na\"{\i}ve second order, has no solutions for the SCSA equations. A solution, however, appears when using a second order in $1/(d-D)$, leading to $\eta=0.789...$ for physical crystalline membranes. The paper ends with a discussion of the meaning of the results. 

\end{section}
\begin{section}{Flat Phase of a Crystalline Membrane} \label{Sec:Flat}
In general dimensionality, i.e., a $D$-dimensional membrane embedded in a $d$-dimensional world, a useful parameterization of a membrane is the Monge representation. Displacements inside the membrane are parametrized using a $D$-dimensional phonon field $\vec{u}$, and the out-of-plane field by a $d_C=d-D$ dimensional field $\bm{h}$. As a result, a particle located on the unperturbed (flat) membrane in coordinate $\vec{x}$, is displaced due to the perturbations to a location $\underline r=(\vec{x}+\vec{u},\bm{h})$ (in $d$ dimensions). 

One assumes an asymptotically flat geometry with small out-of-plane perturbations, such that $\vec{u}$ and $\bm{h}$ are functions of $\vec{x}$. To leading order in field gradients, the resulting free energy is a sum of the membrane's elastic and bending energies: 
\begin{equation}
F[\vec{u},\bm{h}]=\frac{1}{2}\int d^D{\vec{x}} \left[\kappa (\nabla^2 \bm{h})^2 + 2\mu
  u_{ij}u_{ij} + \lambda u_{ii}^2 \right].
\end{equation}
Here, $\kappa$ is the bending energy, $\lambda$ is the first lam\'e constant, $\mu$ is the shear modulus, and $u_{ij}$ is the strain tensor,
defined by $u_{ij} \equiv \frac{1}{2} \left( \partial_i u_j+\partial_j
u_i +\partial_i \bm{h} \cdot \partial_j \bm{h} \right)$, for $i,\,j=1,...,D$.  The in-plane phonon fields
appear only quadratically, thus can be integrated out of the free
energy. The resulting effective energy depends only on $\bm{h}$, and in Fourier space receives the form \cite{1992PRLSCSA}:
\begin{eqnarray} \label{Eq:Elastic_energy_after_integration}
\lefteqn {F_{eff}[\bm{h}]= \frac{1}{2}\int \frac{d^D\vec{q}}{(2\pi)^D} {\Huge\{} \kappa q^4|\bm{h}_{\vec{q}}|^2 +}\\ \nonumber &&+\left. \int \frac{d^D\vec{k}}{(2\pi)^D} \int \frac{d^D\vec{k'}}{(2\pi)^D} \frac{R^{(D)}(\vec{k},\vec{k'},q)}{4(d-D)} \bm{h}_{\vec{k}}\cdot \bm{h}_{\vec{q}-\vec{k}}\bm{h}_{\vec{k'}} \bm{h}_{-\vec{q}-\vec{k'}}\right\}.
\end{eqnarray}
The integrated out elasticity is hidden in the effective four point coupling,
\begin{equation}  \label{Eq:interaction}
\small R^{(D)}(\vec{k},\vec{k}',\vec{q})=2\mu (\vec{k} \mathrm{P^T}(\vec{q}) \vec{k}')^2 +\frac{2\mu\lambda}{2\mu+\lambda}(\vec{k} \mathrm{P^T}(\vec{q}) \vec{k})(\vec{k}' \mathrm{P^T}(\vec{q}) \vec{k}'),
\end{equation}
where the transverse projection operator is defined as
$\mathrm{P^T_{ij}}(\vec{q})=\left(\delta_{ij}-\frac{q_iq_j}{q^2}\right)$. To this end, two main results should be emphasized: 
\begin{itemize}
\item The dependence of the effective interaction on the dimension of the embedding space is trivial, and goes like $1/d_C$. As a result, every interaction of four-$\bm{h}$ fields will contribute a factor of $1/d_C$ to the calculated observable.
\item The $\bm{h}$ field propagator has the property $\langle \bm{h} (\vec{q}) \bm{h} (-\vec{q})\rangle=\mathrm{\bold{I}}_{d_C\times d_C} k_B T G(q)$, where $\mathrm{\bold{I}}_{d_C\times d_C}$ is the identity matrix in $d_C$ dimensions ($k_B$ is Boltzmann constant and $T$ is the temperature). Thus, in the diagrammatic formulation of the theory, any closed loop of $\bm{h}$ fields propagator will contribute a factor of $d_C$ to the diagram. 
\end{itemize}
In view of these points and Eqs.~(\ref{Eq:Elastic_energy_after_integration}~-~\ref{Eq:interaction}), two perturbative procedures to solve the system are apparent. The first is to organize the diagrams in powers of $1/d_C$. Evidently, a diagram with $L_h$ loops and $N_R$ interactions will contribute to the $n$th order in this expansion {\it{iff}} $n=N_R-L_h$. The other perturbative expansion is in the number of loops, which basically implies that the effective four point elastic interaction is small compared to the bending energy. However, this criterion is valid only for length scales smaller than $\sqrt{\kappa^2/k_B TY}$, while we are interested in the long wavelength behavior of the system ($Y$ is some combination of the elastic constants). 

Let us now concentrate in the physical case of $D=2$ and $d=3$. In this case the interaction is completely separable: $R^{(2)}(\vec{k},\vec{k}',\vec{q})=K_0 [\hat{q}\times\vec{k}]^2 [\hat{q}\times\vec{k'}]^2$, where $K_0=\frac{4\mu(\mu+\lambda)}{2\mu+\lambda}$ is the 2D Young
modulus, and $\hat{q}=\vec{q}/q$. The separability allows writing the Feynman rules of Fig.~\ref{Fig:Feynman_rules}, for the $h$ field and for the screening of the interaction. 
\begin{figure}[t]
\rotatebox{0}{\resizebox{4cm}{!}{
\includegraphics[clip=true,viewport=7cm 19cm 14cm 27.cm]{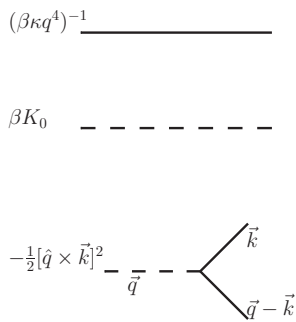}} }
\caption{Feynman rules for the physical theory.}
\label{Fig:Feynman_rules}
\end{figure}
The scale evolution of these fields is governed by the two point diagrams. Thus, we define:  
\begin{equation} \label{Eq:SCSA_dimless}
y(q)=\frac{\kappa_R(q)}{\kappa} \,\,\,\,;\,\,\,\, z(q)=\frac{K_R(q)}{K_0} .
\end{equation}
with $\kappa_R(q)\equiv k_B T (\langle h(q) h(-q) \rangle q^4)^{-1}$, and $K_R(q)$ the screened interaction. This scale evolution can be written in terms of Dyson equations \cite{2009ArxivGazit}:
\begin{eqnarray} \label{Eq:kappa_R}
y(q)& = & 1+\left(\frac{q_0}{q}\right)^2\Sigma(q) 
\\ \label{Eq:K_R}z(q)^{-1} &=&  1+\frac{1}{2}\left(\frac{q_0}{q}\right)^2\Psi(q).
\end{eqnarray}
Here, $q_0=\sqrt{\frac{K_0k_BT}{\kappa^2}}$, $\Sigma(q)$ is the sum of all $1PI$ two-point diagrams, and $\Psi(q)$ is the sum of all $1PI$ four-point diagrams. The latter have a symmetry factor of $\frac{1}{2}$ which is factored out of $\Psi(q)$.
In the long-wavelength limit one assumes a critical behavior for the couplings, $y(q)=y_0(q/q_0)^{-\eta}$ and $z(q)=z_0(q/q_0)^{\eta_u}$, which imply the long wavelength behavior to the $1PI$ amplitudes.
\end{section}
\begin{section} {Self Consistent Screening Approximation}
SCSA is an extension of a consistent perturbative expansion, in which one replaces every propagator with the dressed propagator, and every interaction with the screened interaction, however cutting the series of diagrams taken into account in the calculation of $\Sigma$ and $\Psi$ according to the order of the original expansion. Then, one solves Eqs.~(\ref{Eq:kappa_R}~-~\ref{Eq:K_R}) self-consistently. For example, in the seminal work of  Radzihovsky and Le Doussal \cite{1992PRLSCSA}, they extended a leading order expansion in $1/d_C$ of $\Sigma$ and $\Psi$:
\begin{eqnarray} \label{Eq:Self_Consistent}
 \Sigma(q) &=& \int \frac{d^2\vec{k}}{(2\pi)^2} z(qk)
\frac{[\hat{k} \times \hat{q}]^4}{y(q|\hat{q}-\vec{k}|) |\hat{q}-\vec{k}|^4} \\ \nonumber 
\Psi(q) &=& \int \frac{d^2\vec{k}}{(2\pi)^2}
\frac{[\hat{k} \times \hat{q}]^4}{y(qk) y(q|\hat{q}-\vec{k}|) |\hat{q}-\vec{k}|^4}
\end{eqnarray}
Note that we use an integration variable scaled by $q$, making it easier to analyze the momentum scaling of the theory, and its long wavelength behavior. 
They then solved these equations self-consistently in the long-wavelength limit $q\rightarrow 0$, assuming an algebraic behavior for the couplings, and defining,
\begin{eqnarray}\label{Eq:I_1}
I_1(a,b)&=&\int \frac{d^2\vec{k}}{(2\pi)^2}
k^a\frac{[\hat{k} \times \hat{q}]^4}{ |\hat{q}-\vec{k}|^{4-b}} = \\ \nonumber
&=&\frac{3}{16\pi} \frac{\Gamma(1-\frac{a+b}{2})\Gamma(1+\frac{a}{2})\Gamma(1+\frac{b}{2})}{\Gamma(2+\frac{a+b}{2})\Gamma(2-\frac{a}{2})\Gamma(2-\frac{b}{2})}.
\end{eqnarray}
One notices that the constant in Eqs.~(\ref{Eq:kappa_R}~-~\ref{Eq:K_R}) can be neglected when $q\rightarrow 0$, so the SCSA equations in the long-wavelength limit are:
\begin{eqnarray} 
 y_0\left( \frac{q}{q_0}\right)^{-\eta} &\approx &\left(\frac{q_0}{q}\right)^{2-\eta-\eta_u} z_0y_0^{-1} I_1(\eta_u,\eta)   \nonumber \\ \label{Eq:Self_Consistent_small_q} 
z_0^{-1}\left( \frac{q}{q_0}\right)^{-\eta_u} &\approx & \frac{1}{2}\left(\frac{q_0}{q}\right)^{2-2\eta} y_0^{-2} I_1(\eta,\eta) 
\end{eqnarray}
In order for these equations to be valid for all $q$ (in the relevant regime of $q\ll q_0$), the scaling relation $\eta_u=2-2\eta$ should be fulfilled. Dividing the equations one reaches an equation for $\eta$, which can be solved analytically to give $\eta=\frac{4}{1+\sqrt{15}}$. Radzihovsky and Le Doussal have solved for $\eta$ at general $D$ and $d$, and showed that this approach recovers other approximations, viz large $d$ limit, expansion in small $\epsilon=4-D$, and $d=D$. The latter is unique to the SCSA for crystalline membranes, as it is not the case for the SCSA for ${\cal O}(n)$ model \cite{1974PhysRevLettBray}. 

Albeit these advantages, the SCSA is an uncontrolled approximation, as it includes a partial summation of diagrams mixing different orders. Thus, it is useful to go beyond the one-loop approximation and check deviations from it. 
\end{section}
\begin{section}{Higher Orders contribution to SCSA}

First, let us comment regarding the effect of including higher orders on one of the aforementioned results  of the SCSA -- the prediction of the scaling relation $\eta_u=2-2\eta$. It can be easily demonstrated that this result survives all orders. This can be concluded by analyzing the topology of the diagrams. Let us assume a diagram of specific order. Creating a higher order diagram from this diagram can be achieved in two ways (within SCSA): (i) inserting an interaction connecting two $h$ field propagators (solid lines); (ii) explicitly plotting one of the diagrams resummed in lower SCSA orders, and repeating point (i) for this new diagram. 

Clearly, only point (i) has to be analyzed, as lower SCSA orders are assumed to fulfill the scaling relation. Let us assume that the lines we cut carry momenta $q\vec{k}$ and $q\vec{k}'$. We cut these by an interaction with momentum $q\vec{\tilde{k}}$, over which we integrate (so that the integration measure is $q^{2} d^2\vec{\tilde{k}}$, and the integration variable is dimensionless). Following the Feynman rules, the new diagram will differ from its ancestor by the additional factor:
\begin{equation}
q^{-2}\int \frac{d^2\vec{\tilde{k}}}{(2\pi)^2}\frac{(-\beta K_R(q\tilde{k}))[\hat{\tilde{k}}\times \vec{k}]^2[\hat{\tilde{k}}\times \vec{k}']^2}{\beta \kappa_R(q|\vec{\tilde{k}}+\vec{k}|)|\vec{\tilde{k}}+\vec{k}|^4\beta \kappa_R(q|\vec{\tilde{k}}-\vec{k}'|)|\vec{\tilde{k}}-\vec{k}'|^4}.
\end{equation} 
Taking the long wavelength limit, and moving to the dimensionless elasticity and rigidity:
\begin{eqnarray}
&&-(z_0 y_0^{-2})\left(\frac{q}{q_0}\right)^{-2+\eta_u+2\eta}\cdot \\ \nonumber &\cdot& \int \frac{d^2\vec{\tilde{k}}}{(2\pi)^2}\frac{[\vec{\tilde{k}}\times \vec{k}]^2[\vec{\tilde{k}}\times \vec{k}']^2}{k'^{4-\eta_u}|\vec{\tilde{k}}+\vec{k}|^{4-\eta}|\vec{\tilde{k}}-\vec{k}'|^{4-\eta}}.
\end{eqnarray} 
In order for this addition not to affect the $q$ behavior, one has to demand that the power will vanish, thus reproducing the scaling relation. This can be easily checked to be valid at general dimensionality.

Additional conclusion is that the general form of the SCSA equations at the long wavelength limit are:
\begin{eqnarray} 
1& = & z_0y_0^{-2}\sigma(z_0y_0^{-2};\eta) ,\\ \nonumber
1& = & z_0y_0^{-2}\psi(z_0y_0^{-2};\eta).
\end{eqnarray}
Where $\sigma=\left(\frac{q_0}{q}\right)^{2-\eta}\frac{y_0}{z_0}\Sigma(q)$, and  $\psi=\frac{1}{2}\left(\frac{q_0}{q}\right)^{2-\eta_u}y_0^2\Psi(q)$, are {\it independent} of the momentum $q$, and depend on the universal amplitude $z_0y_0^{-2}$ and critical exponent $\eta$. Moreover, $\sigma$ ($\psi$) is a polynomial in $z_0y_0^{-2}$, whose coefficients have alternating signs and are functions only of $\eta$. The power of the polynomial equals the number of internal interaction lines minus one (internal interaction lines) taken into account in the calculation of $\sigma$ ($\psi$). 

For completeness, though it is a deviation from the physical case $D=2$, we discuss one of the most important results of SCSA, predicting that crystalline membranes are auxetic, i.e., have a negative poisson ratio. In particular SCSA predicts, for all $D\ne\{1,2\}$, $\lim_{q\rightarrow 0} \frac{\lambda(q)}{\mu(q)}=-\frac{2}{D+1}$. Here, we show that this result can be extended to the Dyson equation level, i.e., to all orders of SCSA. The proof follows the lines of Ref.~\cite{1992PRLSCSA}. We start by rewriting the general dimensionality interaction of Eq.~(\ref{Eq:interaction}) as $R^{(D)}(\vec{k},\vec{k'},\vec{q})=k_i k_j k'_l k'_m \rho(q)_{ij,lm}$, with $\tensor\rho(\vec q)=2\mu \tensor M(\vec{q})+2b \tensor N(\vec{q})$, and $b=\mu(2\mu+D\lambda)/(2\mu+\lambda)$. $\tensor{M}$ and $\tensor{N}$ are defined as (assuming $D\ne1$):
\begin{eqnarray}
N_{ij,lm}& = &\frac{1}{D-1}\mathrm{P}^\mathrm{T}_{ij}(\vec{q})\mathrm{P}^\mathrm{T}_{lm}(\vec{q})
\nonumber \\
M_{ij,lm}& = &\frac{1}{2}\left(\mathrm{P}^\mathrm{T}_{il}(\vec{q})\mathrm{P}^\mathrm{T}_{jm}(\vec{q})+\mathrm{P}^\mathrm{T}_{im}(\vec{q})\mathrm{P}^\mathrm{T}_{jl}(\vec{q}) \right) - N_{ij,lm}
\nonumber.
\end{eqnarray}
We note that $\tensor M$ and $\tensor N$ are orthogonal for all $D\ne2$ (since $\tensor M(D=2)=0$).
We proceed by writing a Dyson equation for the interaction, generalizing Eq.~(\ref{Eq:K_R}), as $\tensor \rho_R=\tensor \rho - \tensor \rho \cdot \tensor \Pi\cdot \tensor \rho_R$, where $\tensor \Pi$ is the four point $1PI$ amplitude. Due to the projection operators $\tensor \Pi= \tensor S \pi(q)$, where $\tensor S$ is the totally symmetric tensor $S_{ij,lm}=\delta_{ij,lm}+\delta_{il,jm}+\delta_{im,jl}$. 
Using the orthogonality of $\tensor M$ and $\tensor N$ for $D\ne2$, one can now write a Dyson equation for each of the elastic coefficients $\mu$ and $b$:
\begin{equation}
b_R(q)=\frac{b}{1+(D+1)b\pi(q)}\,\,\,\,\,;\,\,\,\,\,\mu_R(q)=\frac{\mu}{1+2\mu\pi(q)}
\nonumber.
\end{equation}
Taking the long wavelength limit, $\pi(q) \sim q^{-\eta_u}$, one gets the universal result $\lim_{q\rightarrow 0} \frac{\lambda(q)}{\mu(q)}=-\frac{2}{D+1}$ for $D\ne\{1,2\}$. As this result does not depend on the order of the expansion, I conjecture that this is the exact result.

Alas, this proof cannot be extended to $D=2$ due to the separability of the interaction in this dimensionality. Monte-Carlo simulations have shown that this result extends to $D=2$, giving $\nu=-0.32(4)$, compared to the $\nu=-1/3$ SCSA extrapolation \cite{1997EPLFalcioni}, indicating that an analytic continuation of this prediction to $D=2$ is exact. A different result was calculated in Ref.\cite{1996PhysRevEZhang}. There, a molecular dynamics approach led to a long-wavelength Poisson ration of $\nu=-0.16(1)$. Hence, more numerical investigations are needed to settle this difference.

\end{section} 
\begin{section} {SCSA for physical membranes applied to second order}
\begin{figure}[t]
\rotatebox{0}{\resizebox{7.5cm}{!}{
\includegraphics[clip=true,viewport=2.5cm 18.5cm 19cm 27.5cm]{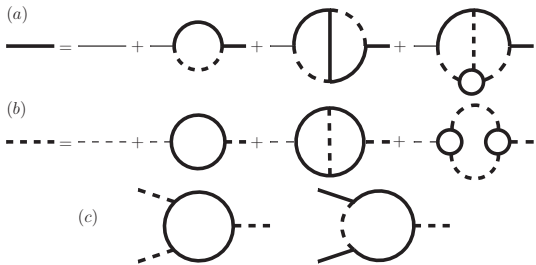}} }
\caption{Diagrammatic form of the self consistent equations for the (a) $h$ field propagator and (b) interaction. A solid line corresponds to the $h$ field propagator, whereas a dashed line corresponds to the interaction. Thick lines are dressed. The equations shown here correspond to the SCSA applied to second order expansion in $1/d_C$. In the first order SCSA \cite{1992PRLSCSA} one takes into account  these equations only up to the one-loop diagram (the first two diagrams in each equation). A na\"{\i}ve two bubble expansion takes into account the first three diagrams in each equation.
(c) Second order interaction vertices.}
\label{Fig:graphic_equations}
\end{figure}

As above mentioned 
extending the calculation to second order is not a fully defined procedure. In this section we will study this issue, and check the exponents implied by each procedure. In Fig.~\ref{Fig:graphic_equations}, the SCSA expansion is presented in a diagrammatic form. If the second order is just a two-bubble expansion, one has to cut these sums and neglect the last diagram in each sum. In order to use a $1/d_C$ expansion, one has to include also those diagrams, which include more than two loops, however satisfying the condition $2=N_R-L_h$. {\it A priori} only the $1/d_C$ expansion is guaranteed to lead to a correction to a lower order theory, as only the $1/d_C$ based SCSA reduces to a controlled method. As will be shown, this fact reveals itself in the case of the flat phase of physical membranes, as no solution is found to the two-bubble expansion.

In an algebraic form, the SCSA equations to order $1/d_C^2$ are ($x=z_0y_0^{-2}$):  
\begin{eqnarray}
\label{Eq:second_order_SCSA}
1& = & a_1(\eta) x - a_2(\eta) x^2 + a_3(\eta) x^3,\\ \nonumber
1& = & b_1(\eta) x - b_2(\eta) x^2 + b_3(\eta) x^3.
\end{eqnarray}
In a two loop expansion one cuts these equations at two coefficients, i.e., taking $a_3=b_3=0$.
To put the coefficients in an algebraic form, it is convenient to define the following integral, representing the second order correction to the interaction vertex, i.e. an internal diagram in which three legs are connected via a loop (see Fig.~\ref{Fig:graphic_equations}(c)):
\begin{equation} \label{Eq:def_I}
I(a,b,c;\vec{k})=\int \frac{d^2\vec{k'}}{(2\pi)^2} \frac{|\vec{k}'\times \vec{k}|^2 |\vec{k}'\times \hat{q}|^2 |(\vec{k}'-\vec{k})\times(\vec{k}'-\hat{q})|^2}{k'^{4-a}|\vec{k}'- \vec{k}|^{4-b}|\vec{k}'- \hat{q}|^{4-c}}
\end{equation}
where for abbreviation we used the notation $\vec{k}=(k,\theta)$, and $\hat{q}=(1,0)$. Evidently, in our case only the integrals with at least two of $(a,b,c)$ equal $\eta$ contribute. Using this definition the coefficients are:
\begin{eqnarray} 
a_1 &=&  I_1(\eta_u,\eta), \nonumber \\
b_1 &=& \frac{1}{2} I_1(\eta,\eta), \nonumber \\
a_2 &=&   \int \frac{d^2\vec{k}}{(2\pi)^2} \frac{[\vec{k}\times\hat{q}]^2}{k^{4-\eta}|\hat{q}-\vec{k}|^{4-\eta_u}} \label{Eq:Coefficients}I(\eta_u,\eta,\eta;\vec{k}), \\
b_2 &=&  \frac{1}{2} \int \frac{d^2\vec{k}}{(2\pi)^2} \frac{[\vec{k}\times\hat{q}]^2}{k^{4-\eta}|\hat{q}-\vec{k}|^{4-\eta}} \label{Eq:Coefficients}I(\eta,\eta_u,\eta;\vec{k}) \nonumber, \\
a_3 &=&  \int \frac{d^2\vec{k}}{(2\pi)^2} \frac{I(\eta,\eta,\eta;\vec{k})I(\eta-2,\eta,\eta;\vec{k})  }{k^{4-\eta_u}|\hat{q}-\vec{k}|^{4-\eta_u}} \label{Eq:Coefficients} \nonumber, \\
b_3 &=& \frac{1}{2}\int \frac{d^2\vec{k}}{(2\pi)^2} \frac{1}{k^{4-\eta_u}|\hat{q}-\vec{k}|^{4-\eta_u}} \label{Eq:Coefficients}I^2(\eta,\eta,\eta;\vec{k})  \nonumber. 
\end{eqnarray}

\end{section}
\begin{section} {Results and Discussion}
To accomplish the calculation of coefficients $a_1,\dots,a_3$ and $b_1,\dots,b_3$, I use a two dimensional Gauss quadrature. The calculation starts with a tabulation of the function $I(a,b,c;\vec{k})$, which takes ${\cal O}(N^4)$ operations, with $N$ being the number of integration points in the range $(0,1)$. All other integrations are ${\cal O}(N^2)$, thus the tabulation determines the length and scaling of the calculation with $N$. The coefficients are calculated for each $\eta$, and are used to solve each of Eqs.~(\ref{Eq:second_order_SCSA}) for $z_0y_0^{-2}$. If the latter are equal, then $\eta$ is a solution to the SCSA equation. In order to reach a result convergent to $10^{-3}$ it is sufficient to use $N=200$ \cite{EPAPS}. 
 
The calculation shows that when considering a two-bubble expansion, one cannot find a solution to the SCSA equations, i.e., Eqs.~(\ref{Eq:second_order_SCSA}) with $a_3=b_3=0$. However, a calculation to second order in $1/d_C$ leads to 
\begin{equation} \nonumber
\eta=0.78922(5), 
\end{equation}
as well as $z_0y_0^{-2}=12.763(5)$. The latter can be put as a universal relation between the bending rigidity and the Young modulus at large wavelengths:
\begin{equation} \nonumber
\lim_{q\rightarrow 0} \frac{1}{q}\sqrt{\frac{k_BT K_R(q)}{\kappa_R^2(q)}} = 3.573(1). 
\end{equation}

When compared to first order SCSA \cite{1992PRLSCSA}, these results give a measure of the accuracy of the SCSA. In first order SCSA, $\eta=\frac{4}{1+\sqrt{15}}=0.82085...$ and $z_0y_0^{-2}=11.2276...$. Evidently, these are close in values to the second order SCSA. In fact, the critical exponent $\eta$ changes only by about $0.03$. Together with the fact that the SCSA for crystalline membranes coincides  with other theoretical estimates in the limits $d_C\rightarrow \infty$, small $\epsilon=4-D$, and $d_C=0$, this is a signature for the relevance of this approximation for the physical case $d=3$, and $D=2$.  

The origins of this success are unclear, and even enhance when considering how essential the description of SCSA as a $1/d_C$ expansion, i.e., the expansion parameter is $1$, as was shown here. One explanation might be hidden in the fact that contrary to SCSA of other field theories, the most famous is the ${\cal O}(n)$ model \cite{1974PhysRevLettBray}, the correction to the interaction vertex is finite, and does not demand any regularization. Additional sources might be hidden in the symmetries of the $D=2$ problem \cite{MouhannaUnPub}. 

The current calculation predicts the other two scaling exponents as well, $\eta_u=2-2\eta=0.4216(1)$, and $\zeta=0.60539(3)$. The latter is in very good agreement with the experimental value $\zeta=0.65(10)$ \cite{1993SciSchmidt}, as well as with the molecular dynamics value $\zeta=0.59(2)$, and goes in the correct direction to the Monte-Carlo simulation result $\zeta=0.64(2)$. The calculated $\eta$ agree, on average, with the different numerical simulations, whose results are in the range $\eta=0.75-0.85$ \cite{1996JPhBowick,1993PhysRevEZhang,2009arXivLos}. The current calculation is also compatible with the non-perturbative RG result $\eta=0.849$ \cite{2009PhysRevEKownacki}, considering the fact that the latter included only low powers in the derivative and field expansion for the action. Stability is found in this approach as well, as the authors find the result stable to changes in the cutoff function. Extending these two non-perturbative approaches to pinpoint the theoretical predictions for the properties of physical crystalline membranes, is highly called for. A rigorous way to estimate the quality of SCSA is by computing the flat phase fixed point to order $\epsilon^2$, and comparing it to the SCSA prediction of the second order $\epsilon$ expansion. The difference between the predictions of these theories, if found, will indicate the level of precision of SCSA \cite{RadzihovskyUnPub}.

It would be interesting to check the current work using experiments and numerical simulations for various interparticle potentials, intended to probe the universality of the membrane theory, not only for the critical exponents, but also for the universal relation predicted to exist between the bending rigidity and the Young modulus at large distances. Of particular interest is to verify whether the exact result for the Poisson ratio at long wavelength that was given here for all membrane dimensionality, excluding $D=1,\,2$, can be extended to these physical cases.
\end{section}
\begin{section}{Acknowledgments}
I thank D. Mouhanna and L. Radzihovsky for valuable discussions and comments. This work was supported by DOE grant number DE-FG02-00ER41132. 
\end{section}

\end{document}